\documentclass[runningheads]{llncs}
\usepackage{amssymb}
\usepackage{bbding} 
\usepackage{graphicx}
\usepackage[hidelinks]{hyperref} 
\renewcommand{\thefootnote}{\fnsymbol{footnote}}

\begin{document}
\title{SmellDoc: Extending Elastic Stack for Microservice Bad Smell Detection and Visualization}

\author{
Yongchao Xing$^{\dagger}$\and
Weipan Yang$^{\dagger}$ \and
Yiming Lv \and
Dianhui Chu \and
Zhiying Tu\Envelope
}

\institute{
Faculty of Computing, Harbin Institute of Technology, China\\
\email{\{22B903085, 23S030135, 24S130271\}@stu.hit.edu.cn,\\
\{chudh, tzy\_hit\}@hit.edu.cn}
}
\titlerunning{SmellDoc: Elastic Stack for Microservice Smell}
\authorrunning{Y. Xing, W. Yang, Y. Lv, D. Chu, and Z. Tu} 

%
%
%
\maketitle              
\footnotetext{$^{\dagger}$ These authors contributed equally to this work.\\
\Envelope\ Corresponding author: tzy\_hit@hit.edu.cn}

\setcounter{footnote}{0}
\renewcommand{\thefootnote}{\arabic{footnote}}
\vspace{-1\baselineskip}
\begin{abstract}
Microservices have become a mainstream architectural paradigm, yet microservice bad smells can significantly harm maintainability and performance. Existing detection tools often produce obscure outputs and lack effective integration with runtime observability, making it difficult for operators to interpret results and take timely action. To address this gap, we propose SmellDoc, a customized framework based on Elastic Stack. SmellDoc extends the native observability dashboard with a microservice bad smell detection plugin, integrating detection, knowledge, and health monitoring. It introduces a Custom-Business-Collector to capture business-level metrics, a Re-integration Collector to aggregate heterogeneous runtime data, and detection components that combine static and runtime analyses. SmellDoc supports a knowledge base of 84 smell types and enables detection of 24 representative smells across architectural, runtime, and performance categories. Results are visualized in Kibana through multiple views, providing operators with actionable insights. Case studies on a benchmark microservice system demonstrate that SmellDoc is effective and usable in detecting, visualizing, and analyzing smells, thus enhancing runtime observability and accelerating troubleshooting to maintain a high level of Quality of Service.
\end{abstract}
\section{Introduction}
With the rise of cloud-native technologies and DevOps, microservices have become a dominant architectural paradigm~\cite{yang2024feature}. However, poor design and development practices, often referred to as microservice bad smells (MBSs), can harm maintainability, scalability, and performance, while increasing the likelihood of defects. Thus, promptly detecting MBSs and gaining a clear understanding of their impact is crucial for timely refactoring and sustaining a high Quality of Service (QoS)~\cite{xing2025saber}.

Existing detection tools (e.g., MARS~\cite{tighilt2023}) mainly output results in unreadable JSON formats, limiting their usefulness. Although some efforts (e.g., MAIG~\cite{gamage2021using}, $\mu$FRESHENER~\cite{soldani2021mutosca}) visualize MBSs in service topology graphs, they provide limited insights into how MBSs affect system runtime. Hence, the challenge lies in integrating MBS detection with runtime observability to support intuitive interpretation and decision-making.

Microservices runtime information spans tracing, data access, and performance metrics, supported by tools such as Prometheus, Elastic Stack, and OpenTelemetry. Prometheus excels in infrastructure monitoring but lacks tracing support; OpenTelemetry standardizes collection but requires external visualization; Elastic Stack\footnote{\href{https://www.elastic.co/elastic-stack}{https://www.elastic.co/elastic-stack}} provides a unified solution with tracing, dependency analysis, and built-in visualization through Kibana, which is extensible via plugins. Leveraging this, we propose \textbf{SmellDoc}, a customized Kibana plugin that integrates MBS detection, knowledge, and health visualization into the Elastic observability dashboard. SmellDoc enables operators to correlate MBSs with runtime metrics, accelerating troubleshooting and improving QoS recovery.

The main contributions are as follows:
\vspace{-0.2\baselineskip}

\begin{enumerate}
  \item We design and implement \textbf{SmellDoc}, a Kibana plugin extending Elastic Stack with integrated MBS detection, knowledge, and system health monitoring.  
  \item SmellDoc incorporates a knowledge base of 84 MBS types and supports multiple detection algorithms (static and runtime) using fine-grained service features collected by custom components.  
  \item Case studies on a benchmark system demonstrate the effectiveness and usability of SmellDoc in detecting and visualizing MBSs.  
\end{enumerate}

The rest of this paper is organized as follows: Section 2 presents SmellDoc’s design and implementation. Section 3 validates its effectiveness. Section 4 concludes the paper.
\vspace{-1\baselineskip}
\section{Methodology}
\begin{figure}
\vspace{-1\baselineskip}
\centering
\includegraphics[width=0.7\textwidth]{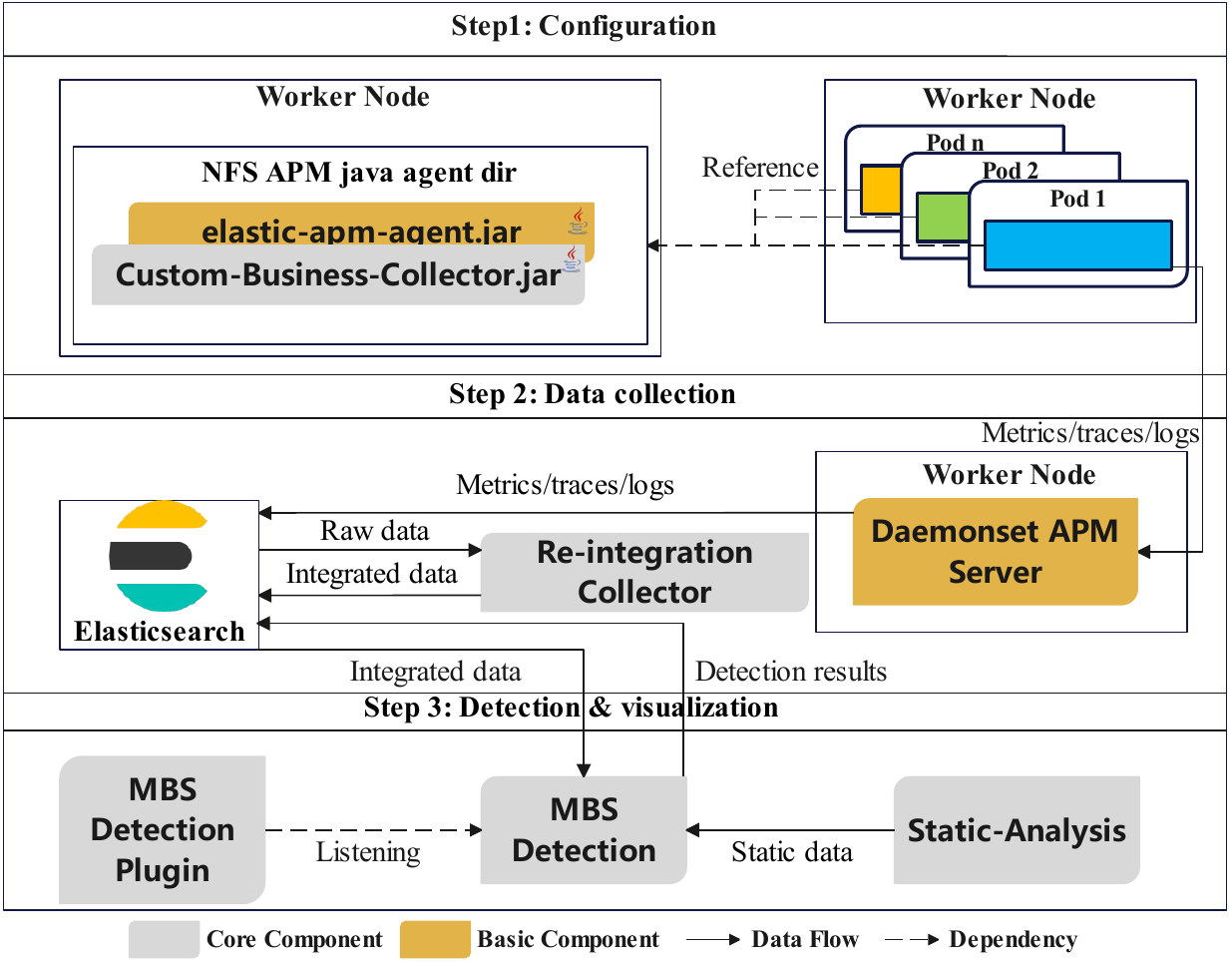}
\caption{The overview of SmellDoc} \label{fig1}
\vspace{-1\baselineskip}
\end{figure}
SmellDoc\footnote{\label{fn:smelldoc}\href{https://github.com/yang66-hash/BSDProject}{https://github.com/yang66-hash/BSDProject}} 
consists of three steps (Fig.~\ref{fig1}). 
\textbf{Step~1 (Configuration):} besides the native \texttt{elastic-apm-agent}, a custom agent 
(\emph{Custom-Business-Collector}, CBC) is introduced to capture business-level metrics. 
When microservices are deployed in Kubernetes, both agents start with the service to collect 
runtime traces and metrics for the APM Server.  
\textbf{Step~2 (Data Collection):} the \emph{Re-integration Collector} (RIC) filters redundant 
records, aggregates multi-source data, and stores them in Elasticsearch.  
\textbf{Step~3 (Detection \& Visualization):} the MBS Detection Component combines static metrics produced by a Static-Analysis component derived from the MBST framework~\cite{xing2025mbst} with runtime data to detect 24 smell types.
Results are stored in Elasticsearch and visualized through the MBSD Plugin, which supports 
interactive exploration of smells alongside native Elastic metrics and traces.

\subsection{Configuration}
SmellDoc extends the native \texttt{elastic-apm-agent} with the CBC, implemented via the Elastic 
APM SDK and Byte Buddy to intercept Spring service methods. CBC records invocation metrics with 
Micrometer and reports them to the APM Server, enabling fine-grained business-level monitoring 
beyond system-level runtime data. Together, the APM Agent and CBC provide comprehensive observability.

\subsection{Data Collection}
The RIC periodically retrieves runtime data from Elasticsearch and integrates three categories: 
(i) external metrics (trace links, SQL calls, cross-service dependencies), 
(ii) internal metrics (JVM-level CPU, memory, GC), and 
(iii) business metrics (service-level invocation frequencies from CBC). 
Compared with Elastic APM’s default support, RIC incrementally introduces business-level counters 
and unified indexing to normalize heterogeneous data before storing them back into Elasticsearch, 
thus improving support for smell detection and health assessment.

\subsection{Detection \& Visualization}
Our pipeline combines a \emph{Static-Analysis Component} with the \emph{MBSD Component}. 
Static analysis extracts service metadata, topology, and fine-grained features, then applies
rule-based checks to detect 12 architecture-level smells (e.g., \emph{ESB Usage},
\emph{Microservice Greedy}). Results and static metrics are fed to MBSD, which augments them
with \emph{runtime data} from Elasticsearch to detect 12 runtime smells. In total, MBSD
produces detection results for \textbf{24 smell types}, all stored in Elasticsearch
(definitions and algorithms in our repository\footnote{\href{https://github.com/yang66-hash/BSDProject/blob/master/README.md}{https://github.com/yang66-hash/BSDProject/blob/master/README.md}}).

For usability, we integrate the MBSD Plugin into the \emph{Elastic Observability}
dashboard with three tabs: (1) \textbf{Bad Smell Knowledges Tab}:\textbf{Bad Smell Knowledges}: standardized definitions and historical records by smell type; (2) \textbf{Monitor Tab}: \textbf{Monitor}: system-wide views for Kubernetes deployments with service/instance KPIs; (3) \textbf{Detection Tab}:\textbf{Detection}: current distributions and statistics, algorithm online/offline status, and visualized trends; per-service histories are available as charts and JSON. This integration couples smell detection with runtime monitoring, enabling timely decisions
and mitigating the accumulation of harmful smells.
\vspace{-1\baselineskip}
\section{Case Study}
\vspace{-1\baselineskip}
We select the open-source microservice system Property Management Cloud\footnote{\href{https://github.com/yang66-hash/PropertyManagementCloud}{https://github.com/yang66-hash/PropertyManagementCloud}} as case. The system is deployed on a Kubernetes (K8s) cluster together with all components of the SmellDoc framework to validate its effectiveness for microservice smell detection and visualization. Property Management Cloud is implemented with Spring Cloud and comprises five microservice modules. Once the services are stably running in the cluster, users can open the Detection tab to view the list of monitored microservices and a system-level overview of MBSs detection.
\vspace{-1.5\baselineskip}
\begin{figure}[!ht]
\centering
\includegraphics[width=0.7\textwidth]{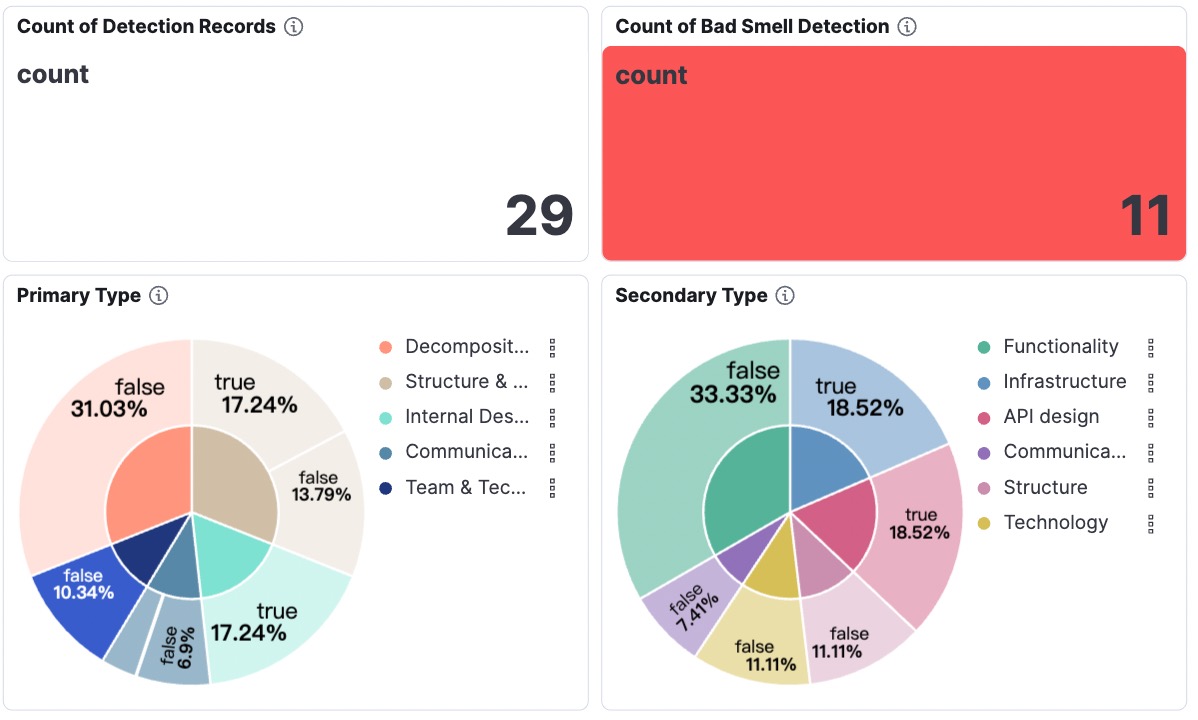}
\caption{Detection card} \label{fig2}
\vspace{-1\baselineskip}
\end{figure}
Fig.~\ref{fig2} presents an example of the detection card. In this example, the system has executed 29 MBSs detections, among which 11 detections reported MBSs. The two-ring pie chart in Fig.~\ref{fig2} summarizes MBS-type distributions: the inner ring shows the proportion of MBS types by Primary Type (top-level category in the MBSs knowledge base) and Secondary Type (second-level category), while the outer ring shows the proportions of detected (true) vs. not detected (false) MBS types. This view allows users to quickly identify the categories to which the currently detected MBSs belong, enabling targeted remediation.

\begin{figure}
\vspace{-1\baselineskip}
\centering
\includegraphics[width=\textwidth]{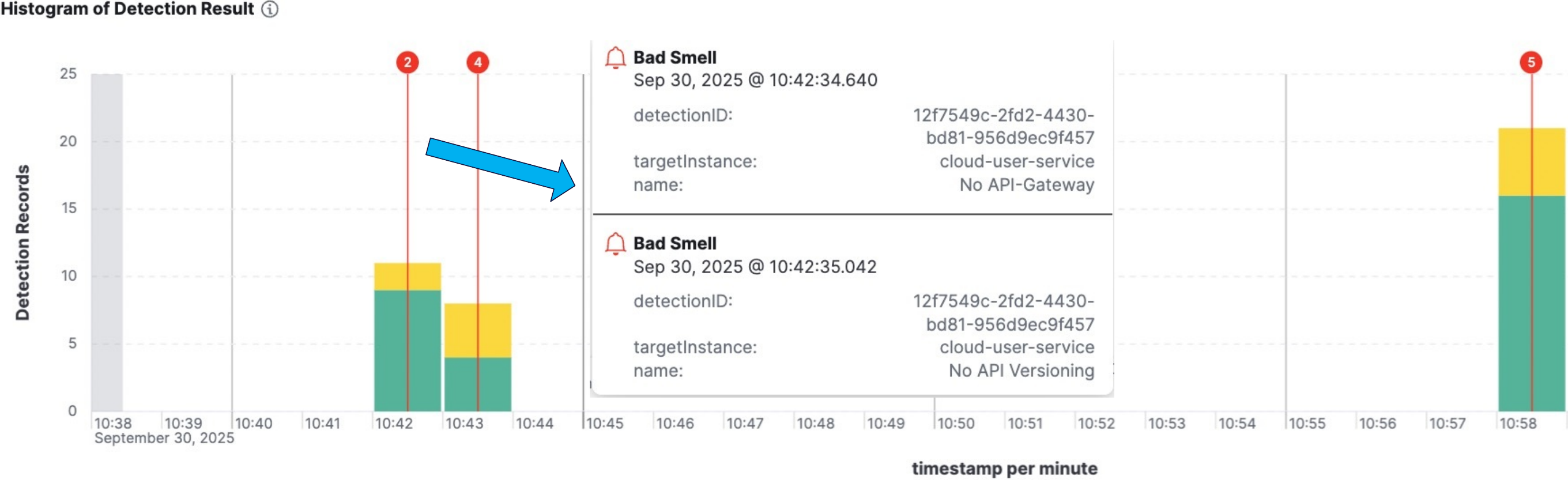}
\caption{History results} \label{fig3}
\end{figure}

Fig.~\ref{fig3} depicts the history of detection results. Smells were observed in three time windows. For example, in the first time window, No API Versioning and No API-Gateway smells were detected in the cloud-user-service. A complete demonstration video using Property Management Cloud is available at \url{https://www.youtube.com/watch?v=a8WBSHTj4J0&t=3s}. The video walks through the operational workflow of SmellDoc and the usage of the MBSD plugin tabs.
\vspace{-3\baselineskip}
\section{Conclusion}
We introduced SmellDoc, an Elastic Stack extension for detecting and visualizing MBSs. Combining static and runtime analyses with fine-grained metrics, it supports 24 smell types and integrates results into Kibana dashboards. Case studies on a system confirmed its effectiveness and usability. SmellDoc shows that embedding smell detection into observability can enhance monitoring, speed troubleshooting, and help sustain high QoS in microservice systems.
\bibliographystyle{splncs04}
\bibliography{springer-bibliography}
\end{document}